\documentclass[12pt,preprint]{aastex}

\shorttitle{Spatially Resolved [FeII] in NGC 5135}
\shortauthors{Colina et al.}

\begin{document}

\title{Spatially Resolved [FeII] 1.64$\mu$m Emission in NGC 5135. \\
    Clues for Understanding the Origin of the Hard X-rays in Luminous Infrared Galaxies}

\author{L. Colina\altaffilmark{1}, M. Pereira-Santaella\altaffilmark{1}, A. Alonso-Herrero\altaffilmark{1,2}, A.G. Bedregal\altaffilmark{3}, S. Arribas\altaffilmark{1} }
\email{colina@cab.inta-csic.es}

\altaffiltext{1}{Astrophysics Department, Center for Astrobiology (CSIC-INTA),
    Torrejon de Ardoz, 28850 Madrid, Spain}
\altaffiltext{2}{Instituto de F\'{\i}sica de Cantabria, CSIC-UC. 39005 Santander, Spain}
\altaffiltext{3}{Department of Astronomy, University of Minnesota, 116 Church St. S.E., Minneapolis, MN 55455, USA}

\begin{abstract}

Spatially resolved near-IR and X-ray imaging of the central region of the Luminous Infrared Galaxy NGC 5135
is presented. The kinematical signatures of strong outflows are detected in the 
[FeII]1.64 $\mu$m emission line in a compact region at 0.9 kpc from the nucleus. 
The derived mechanical energy release is consistent with a supernova rate of 0.05$-$0.1 yr$^{-1}$.  
The apex of the outflowing
gas spatially coincides with the strongest [FeII] emission peak and with the dominant component of
the extranuclear hard X-ray emission.  All these features provide evidence for a plausible
direct physical link between supernova-driven outflows and the hard X-ray emitting gas in a LIRG. This result is
consistent with model predictions of starbursts concentrated in small volumes and with high thermalization efficiencies.
A single high-mass X-ray binary (HMXB) as the major source of the hard X-ray emission although 
not favoured, cannot be ruled out. 
Outside the AGN, the hard X-ray emission in NGC 5135 appears to be dominated by the 
hot ISM produced by supernova explosions in a compact star-forming region, and not by the emission due to HMXB.  
If this scenario is common to U/LIRGs, the hard X-rays would only trace the most compact ($\leq$ 100 pc) regions with high supernova
and star formation densities, therefore a lower limit to their integrated star formation.
The SFR derived in NGC 5135 based on its hard X-ray luminosity is a factor of two and four 
lower than the values obtained from the 24 $\mu$m and soft X-ray luminosities, respectively.

\end{abstract}


\section{Introduction}

Luminous infrared galaxies (LIRGs) with infrared 
(8$-$1000 $\mu$m) luminosities (L$_{IR}$) of  10$^{11}$ L$_{\odot}$ to 10$^{12}$ L$_{\odot}$, are powered mainly 
by star formation with some contribution from an AGN in a fraction of them (Genzel et al. 1998; Nardini et al. 2008;
Alonso-Herrero et al. 2012).  
Aside from
the contribution by the compact AGN, if present, the X-ray luminosity in LIRGs is produced by the star 
formation. As such, correlations between the Star Formation Rate (SFR), independently
derived from the IR luminosity, and the X-ray luminosity are expected. These correlations for both
soft (0.5$-$2 keV) and hard (2$-$10 keV) X-rays have recently been obtained in a sample of LIRGs
(Pereira-Santaella et al. 2011 and references therein). These galaxies follow the same linear correlations found for normal star-forming
galaxies within a factor of 2 scatter.  The physical origin of the extension of these correlations 
to luminous and compact starbursts with SFR of up to 100 M$_{\odot}$ yr$^{-1}$ is not fully clear. While the soft X-ray emission is assumed to be dominated by the diffuse, extended (kpc-size) emission of a hot (10$^6 - 10^7$ K) starburst-driven gas (Strickland et al. 2002; Ranalli et al. 2003; McDowell et al. 2003), the hard X-rays are assumed to be predominantly 
due to the collective emission of luminous high-mass X-ray binaries (HMXB), and therefore associated to unresolved sources (Grimm et al. 2003; Persic et al. 2004; Strickland \& Heckman 2007). However, high angular X-ray imaging of M82 (Strickland \& Heckman 2007) has shown the presence of a diffuse component contributing  about one third of the 3-10 keV luminosity. Moreover, non-AGN ULIRGs and high-luminosity LIRGs show in their integrated spectrum strong Fe XXV
at 6.7 keV, which is not produced by HMXBs but rather likely due to a high-temperature (T $\sim$ 10$^8$ K) thermal gas
(Iwasawa et al. 2009). These authors conclude that unlike for local star-forming galaxies, HMXBs are not the primary source of the hard X-ray band emission seen in non-AGN luminous LIRGs (L$_{IR}$ $\geq 5 \times 10^{11}$ L$_{\odot}$) , suggesting that a transition in the nature of the dominant hard X-ray source in LIRGs occurs somewhere in the range log L$_{IR}$ =11.0-11.73 $L_{\odot}$ (Iwasawa et al. 2011).
These results seem to give support to recent starburst-driven wind models (Strickland \& Heckman 2009) where a very hot plasma 
(10$^7$ K $<$ T $<$ 10$^8$ K) produced by the thermalization of the mechanic energy released by supernova explosions in compact starbursts would
be a major contributor to the hard X-ray emission. A contribution to the Fe XXV emission at 6.7 keV in U/LIRGs could also come from heavily-obscured AGNs.
However, XMM surveys of Seyferts have shown that the 6.7 keV FeXXV line is detected in only a small fraction (3 out of 26) of the surveyed galaxies (Nandra et al.
2007). In addition, recent studies of the nearby edge-on starburst galaxy NGC 253 have demonstrated that the FeXXV 6.7 keV emission is not dominated by the obscured AGN but by an extended hot gas emission outside the nucleus. The origin of this emission is consistent with the presence of supernovae and supernovae remnants, with a minor contribution from HMXBs and cataclysmic variables (Mitsuishi et al. 2011).  

In order to investigate the different scenarios and physical origin of the hard X-ray emission, i.e. extended hot plasma gas associated with 
supernova versus point sources (HMXB, AGN), spatially resolved imaging of supernovae tracers (e.g. radio or near-IR [FeII]1.64$\mu$m emission) and
of the  X-ray emission are needed. So far, such studies have been carried out for local star-forming galaxies such as NGC 253 and M82 (Weaver et al. 2002; Strickland \& Heckman 2007), but no adequate data have been available for more powerful starbursts such LIRGs.    

NGC 5135 is a nearby LIRG (log(L$_{IR}$/L$_{\odot}$) = 11.3, Pereira-Santaella et al. 2011) 
with a Compton-thick AGN  in its nucleus (Levenson et al. 2004), and several 
circumnuclear star-forming regions at distances of one kpc away from the nucleus, identified
by their UV stellar light (Gonzalez-Delgado et al. 1998), (partially)ionized emitting gas (Alonso-Herrero et al. 2006a, Bedregal et al. 2009), and 
mid-IR 8$\mu$m emission (Alonso-Herrero
et al. 2006b). Due to its distance to us (59.2 Mpc), the available near-IR integral field spectroscopy
(Bedregal et al. 2009) and X-ray {\it Chandra} imaging (Levenson et al. 2004) allow us to spatially
resolve the structure of its central few kpc region on scales of hundred of parsecs (1 arcsec corresponds to 
287 pc at the assumed distance).  The contribution of the different energy sources to the hot X-ray emitting gas, 
and the causal connection between the X-ray emission, in particular hard X-ray, and supernova explosions can therefore
be investigated in detail.

\section{Observations and Data}

Near-IR data were obtained with SINFONI, the VLT near-IR integral field spectrograph.
The galaxy was observed in the H and K bands separately with a spectral resolution of
$\sim$ 3000 and $\sim$ 4000, respectively. The configuration during the observations was the
standard seeing-limited, with a scale of 250 mas per spaxel and field-of-view of
$8" \times 8"$. Further details about the calibration process, data reduction and emission line 
measurements can be found elsewhere (Bedregal et al. 2009)

NGC 5135 was observed in X-rays with {\it Chandra} Advanced CCD Imaging Spectrometer (ACIS)
for a total exposure time of 29.3 ks (Levenson et al. 2004). These data have been retrieved from the \textit{Chandra} archive. The \texttt{chandra\_repro} script (CIAO v4.3) was used to reprocess the level 1 data using the latest available calibration (CALDB v4.4.2). This script performs the recommended data reduction steps\footnote{http://cxc.harvard.edu/ciao/threads/data.html} (which includes bad pixel removal, corrections for the coordinates and energy of the events, corrections for charge transfer inefficiency, etc.). The local background was measured in a nearby source-free circular region of radius 50 arcsec located 2 arcmin away from the galaxy nucleus. 
The background lightcurve was checked to filter out high-background intervals. However no flares were detected during the observation.
Only the events with energies in the interval from 0.3 to 10\,keV are considered in the analysis\footnote{The 0.3--10\,keV\ range is the ACIS calibrated energy range.}.
The spectra of the two bright compact sources (regions A and E in Figure \ref{fig1}) were extracted using a circular aperture of radius 1.5\,arcsec.
The spectrum of the B+C region was extracted using a 2$\times$3\,arcsec elliptical aperture.
The response matrix files (RMFs) and the ancillary response files (ARFs) were created assuming a pointlike source.
The spectra were grouped to have at least 20 counts in each spectral bin.

\section{Results}

\subsection{Near-IR Emission Line Imaging}

The spatially resolved [FeII] 1.64$\mu$m, Br$\gamma$ 2.17$\mu$m, and [SiVI] 1.96$\mu$m emission in the central 1.7 kpc by 1.7 kpc
region of NGC 5135 is presented together with the soft- (0.5$-$2.0 kV) and hard (2.0$-$8 keV) X-ray 
emission in Figure 1. Also given is the velocity field of the interstellar medium as traced by the [FeII] line. The near-IR emission lines and the X-ray continuum trace different phases of the (circum)nuclear interstellar 
medium (ISM), showing clear substructures on scales of hundreds of parsecs, and a number of 
characteristics that trace in detail the connection between the existing energy sources (AGN, young stars, shocks), and 
their impact in the ionization and kinematics of the ISM.

\begin{figure*}
\center
\includegraphics[angle=90,scale=0.80]{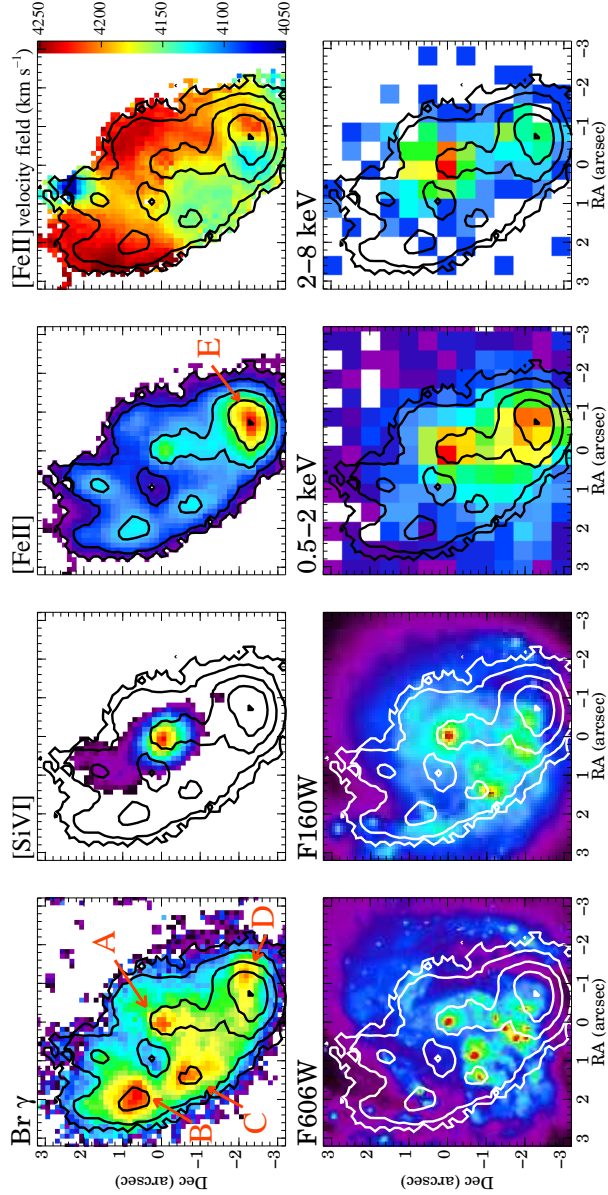}
\caption{Panel presenting the structure of the central region of NGC 5135 according to different near-IR emission lines (Br$\gamma$, [SiVI] 
1.96$\mu$m, [FeII] 1.64$\mu$m), soft (0.5-2 keV), and hard (2-8 keV) X-ray emission, as well as the velocity field of the partially-ionized
[FeII] emitting gas.  The optical (HST WFPC2 F606W filter), and near-IR (HST NICMOS F160W filter)  stellar light distribution 
are also shown for a direct comparison with the structure of the near-IR and X-ray emitting gas. Individual regions like the nucleus (A) and circumnuclear
star-forming regions (B to E) are labeled according to Bedregal et al. (2009). Contours represent the [FeII] emission distribution. At the assumed distance, one arcsec corresponds to 287 pc. \label{fig1}}
\end{figure*}

The AGN (region A in Figure 1) is identified as a strong source of coronal [SiVI] emission and appears as a weak Br$\gamma$ 
and [FeII] emitting source. Several Br$\gamma$ and [FeII] emitting regions (B to E in Figure 1) trace 
circumnuclear star-forming clumps at distances of about 1 kpc from the nucleus. While the brightest 
Br$\gamma$ region (B in Figure 1) shows a low ($\sim$ 2) [FeII]/Br$\gamma$ ratio, the
reverse is true for the brightest [FeII] region (E in Figure 1) where a [FeII]/Br$\gamma$ ratio of $\sim$ 20 is
measured (Bedregal et al. 2009). These different line ratios can be attributed to a relative age difference of about 4 million 
years between the star clusters in both regions. Region B would be the youngest ($\leq$ 6 Myr) where massive main sequence stars are
still producing a large number of ionizing photons, e.g. strong hydrogen recombination lines. On the other hand,
region E would be more evolved and going through the peak of the supernova phase (i.e. 8-12 Myr), therefore producing 
locally a strong [FeII] emission peak, which is spatially coincident with the peak emission at radio wavelengths (see Bedregal et al.
2009 for a discussion on the uncertainties and assumptions). The velocity field of the [FeII] emission around region E shows locally a velocity 
gradient along the east-west
direction, characterized by a projected peak-to-peak velocity of 90 km s$^{-1}$ and an overall projected size of 300 pc.
This structure resembles a biconical outflow with the apex coincident with the peak of the [FeII] emission (see Figure 1), and 
is interpreted as a starburst-driven outflow due to the combined effect of several supernovae in region E, consistent with the high 
supernova explosion rate derived for this region (Bedregal et al. 2009). The 
morphology of the star-forming regions traced by the ionized gas is consistent with a disk inclined by 
about 50 degrees with respect to our line of sight. If the axis of the outflow were oriented perpendicular the plane of the disk, 
the intrinsic size and outflowing velocities of the [FeII] emitting gas would be a factor of 1.5-2 higher than the observed values.

\subsection{X-ray Imaging}

The structure of the X-ray emitting gas is characterized by the presence of two bright soft (0.5-2 keV) and hard (2-8 keV) X-rays
unresolved regions, as well as a prominent diffuse emission component in soft X-rays (see Figure 1, also Levenson et al. 2004). This structure 
presents interesting features when compared with the near-IR line emitting gas (Figure 1 and Table 1 for specific luminosities). 
The brightest soft and hard X-ray emission peaks are associated with the AGN nucleus and therefore coincident with 
the [SiVI] emission peak. The secondary X-ray emission peak is spatially coincident with the 
[FeII] emission peak (region E). Finally, the diffuse soft X-ray emission extends northeast from region E towards region C (secondary 
extranuclear Br$\gamma$ peak) avoiding the brightest Br$\gamma$ emitting region 
(B in Figure 1) and having an overall structure in surface brightness in better agreement to that shown in [FeII] emission.  Moreover,
there appears to be  also a kinematic connection between the X-ray and [FeII] emitting gas as the secondary X-ray peak coincides (within 
the astrometric uncertainties) with the apex of the biconical [FeII] outflow. This strongly supports a causal connection between 
the extremely hot X-ray emitting gas and the supernova explosions in this region of the galaxy (see discussion in section
4.1).

\begin{deluxetable}{ccccc}
\tabletypesize{\scriptsize}
\rotate
\tablecaption{X-ray and Near-IR Line Luminosities for the Nucleus and Circumnuclear Star-forming Regions of NGC 5135}
\tablewidth{0pt}
\tablehead{
\colhead{Region}\tablenotemark{a} & \colhead{L(Br$\gamma$)} & \colhead{L([FeII])} & \colhead{L(0.5-2 keV)} & \colhead{L(2-8 keV)}\\
& 10$^{39}$ erg s$^{-1}$ & 10$^{39}$ erg s$^{-1}$ & 10$^{40}$ erg s$^{-1}$ & 10$^{40}$ erg s$^{-1}$ }
\startdata
A &   0.64 & 3.65 & 4.3 & 13.1\\
B &   0.55 & 1.17 & $--$ & $--$ \\
C &   0.51 & 1.28 & $--$ & $--$ \\
B+C&  1.06 & 2.45 & 1.2 & $<$ 0.5 \\
D &   0.35 & 2.32 & $--$ & $--$ \\
E &   0.30 & 5.50 & $--$ & $--$ \\
D+E & 0.65 & 7.82 & 6.3 & 1.4\tablenotemark{b} \\
\enddata
\tablenotetext{a}{{L}uminosities for the individual regions (A to E) correspond to apertures with diameters 340 pc (near-IR lines), and
860 pc (X-rays). Regions A to E appear as point-like. The quoted diameters correspond to the different angular size of the PSF of 
SINFONI and {\it Chandra}. Region B+C corresponds to an area of 1.2 kpc $\times$ 0.6 kpc as measured in soft X-rays. The X-ray luminosity for
regions D \& E can not be measured independently due to the lack of angular resolution in the {\it Chandra} images.}
\tablenotetext{b}{ Detected at 4.5$\sigma$ level}
\end{deluxetable}

\section{Discussion}

\subsection{The X-ray Emission. Origin of the Spatially Resolved Components} 

As mentioned above, the soft and hard X-ray images show bright nuclear emission (region A in Figure \ref{fig1}), a secondary extranuclear emission coincident with the brightest
[FeII] peak (region E), and a fainter and more diffuse emission, only detectable in soft X-rays, likely associated with the secondary Br$\gamma$ emission
(region C).  The physical mechanisms involved in the emission of the X-rays appears to be different in these regions.
The X-ray spectrum of the nuclear region  has been analyzed in detail by Levenson et al. (2004). They used a model consisting of an absorbed power-law, two thermal plasma components, and two Gaussian emission lines at 6.4 and 1.8\,kV, respectively. The flat power-law ($\Gamma = 0$) and the large equivalent width of the Fe K$\alpha$ emission line (2.4\,keV) indicate that the AGN is highly obscured, with an absorbing column density ($N_{\rm H}$) above 10$^{24}~cm^{-2}$. This conclusion has been recently confirmed by a spectral study of the broad {\it Suzaku} 0.5 $-$ 50 kV range, establishing a columm density $N_{\rm H}$ of $\sim$ 2.5 $\times$ 10$^{24}$~cm$^{-2}$ towards the AGN (Singh et a. 2011)
The spectrum of the B+C region was not explicitly discussed by Levenson et al. (2004). This region is included in their D1 region. Due to the low number of counts of this spectrum (81 counts), Levenson et al. used a simple model with an absorbed power-law. The power-law index is 3$^{+3}_{-1}$ and the $N_{\rm H}$ is lower than 5$\times 10^{21}$\,cm$^{-2}$.

Levenson et al. (2004) also discussed the spectrum of the star-forming region identified here as region E. Their model includes a thermal component at 0.7\,keV, similar to that found in starbursts (see Persic \& Rephaeli 2002), and a power-law ($\Gamma = 2.6$). The Fe abundance was allowed to vary to account for the Fe underabundance with respect to $\alpha$ elements observed in star-forming galaxies (e.g. Strickland et al. 2004; Grimes et al. 2005; Iwasawa et al. 2011; Pereira-Santaella et al. 2011). We repeated the analysis and found that this model provides a good fit to the data ($\chi^2$\slash dof $=$ 34\slash 37, see Table 2). The non-thermal component of the model would be interpreted as
emission associated to high-mass X-ray binaries (HMXB). Although the power-law is steeper than in individual binary systems it might be produced by the combination of several individual sources (Levenson et al. 2004).
Alternatively the spectrum of the region E can be fitted equally well with two thermal components. This model also provides a good fit ($\chi^2$\slash dof $=$ 35\slash 37, see Table 2).
The temperature of the soft thermal component is 0.7\,keV, in agreement with previous Levenson et al. (2004) results. Although the hard thermal component is required (with an F-test significance $>98\%$), its temperature is not well constrained ($kT>2.4$\,keV). This is due to the low signal to noise ratio in the hard X-ray range (only 10 counts are detected above 4\,keV in this region). 
Similar to the nuclear starburst in M82, the soft thermal component would be associated with the extended, diffuse X-ray emitting gas while the hard thermal component would represent extremely hot gas originated in the inner compact region of the starburst directly associated with the strong winds produced by the supernovae (Strickland \& Heckman 2007, 2009).

\begin{figure*}[t]
\includegraphics[scale=0.75]{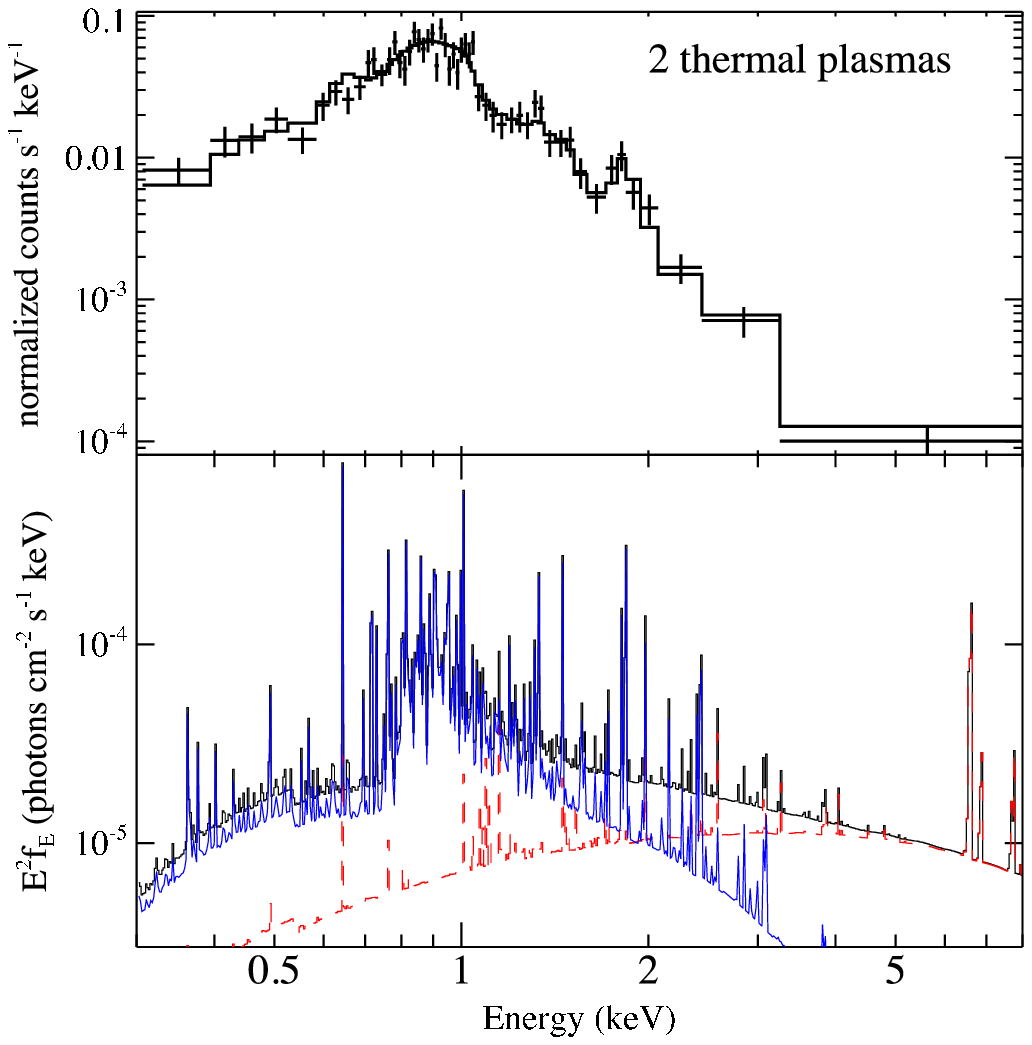}
\includegraphics[scale=0.75]{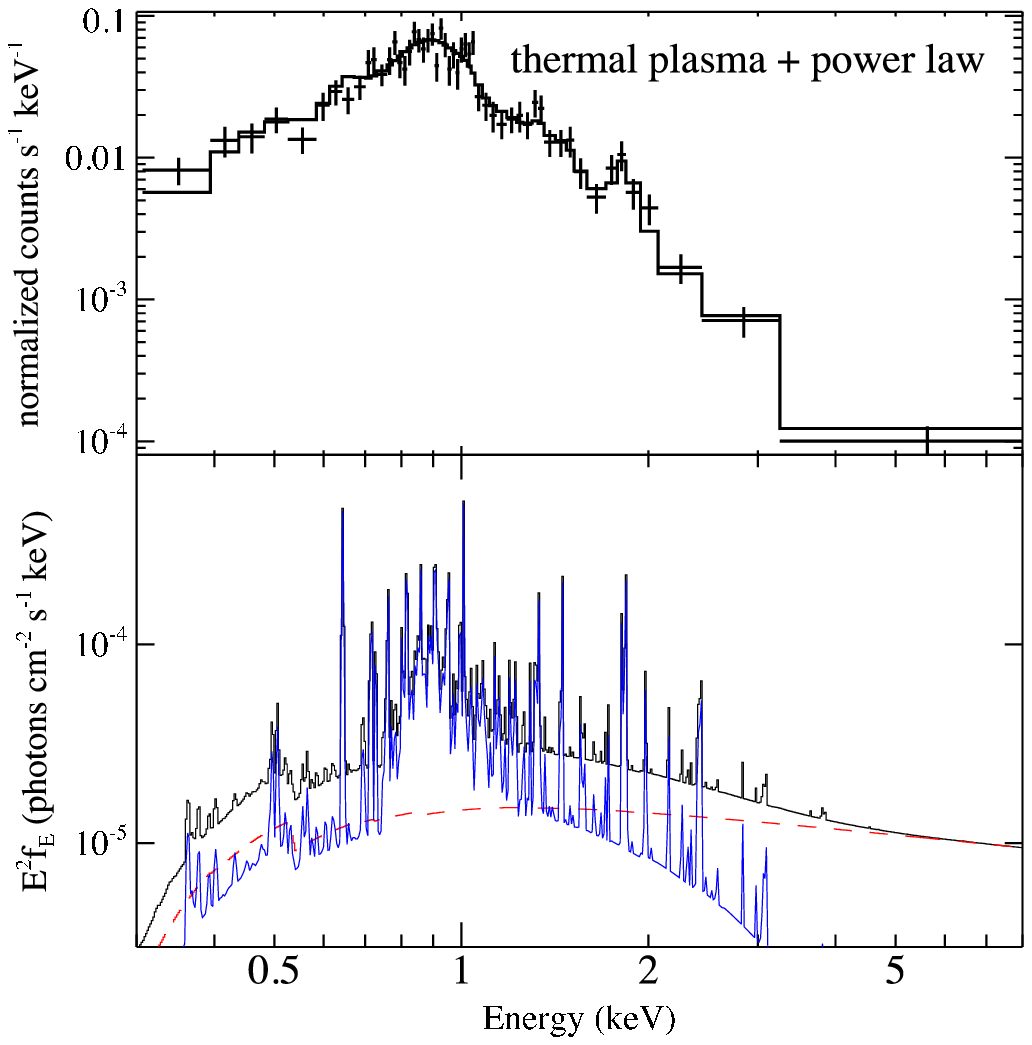}
\caption{Comparison of the two spectral models for the \textit{Chandra} ACIS spectrum of the region E of NGC~5135 (note that due to lack of angular resolution the spectrum includes also region D identified in the near-IR images). The top panel shows the observed data together with the model (solid black line). The panel below shows the spectral components of the model: the soft ($kT\sim 0.7$\,keV) thermal plasma (solid blue line) and the hard component (hot thermal plasma or power-law; dashed red line). The total model is shown in black. The analysis of the X-ray spectral data was done using \texttt{XSPEC} (Arnaud 1996). To account for the Galactic hydrogen column density at the position of NGC~5135 ($N_{\rm H}=$ 4.9 $\times$10$^{20}$\,cm$^{-2}$, Kalberla et al. 2005) the TBabs model (Wilms et al. 2000) was adopted.
\label{fig2}}
\end{figure*}

Figure \ref{fig2} shows the comparison of the two models while the model parameters and goodness of the fit can be found in Table 2.
From the available X-ray spectrum alone both models, and both interpretations on the origin of the hard X-ray emission, are plausible. However the high supernova rate and kinematic evidence of large scale outflows derived from the SINFONI near-IR data for this region favors the second model, i.e. two thermal gas components, as opposed to the HMXB interpretation (see discussion in following 
sections). 
The detection of the ionized Fe K$\alpha$ line at 6.7\,keV would provide a strong direct evidence for the presence of a extremely hot gas component associated to supernovae in this region.

\begin{deluxetable}{lcccccccc}
\tabletypesize{\scriptsize}
\tablecaption{Model parameters for the best fits to the X-ray spectrum of region E}
\tablewidth{0pt}
\tablehead{
\colhead{Model} & \colhead{$N_{\rm H}$} & \colhead{$\Gamma$} & \colhead{$kT_1$} & \colhead{Fe/O\tablenotemark{a}}& \colhead{$kT_2$} & \colhead{$\chi^2$\slash dof} \\
& (10$^{20}$\,cm$^{-2}$) & & (keV) & & (keV)
}
\startdata
plasma + power law\tablenotemark{b} & 8.6$^{+5.4}_{-6.0}$ & 2.5$^{+0.8}_{-0.9}$ & 0.68$^{+0.07}_{-0.07}$ & 0.4$^{+0.2}_{-0.1}$ & \nodata & 34\slash 37\\
Two plasmas\tablenotemark{b} & $<$10 & \nodata & 0.70$^{+0.07}_{-0.06}$& 0.3$^{+0.1}_{-0.1}$ & $>$2.4 & 35\slash 37\\
\enddata
\tablecomments{The Galactic absorption at the position of NGC~5135 is $N_{\rm H, Gal} =$ 4.9$\times$10$^{20}$\,cm$^{-2}$ (Kalberla et al. 2005).}
\tablenotetext{a}{Fe/O abundande ratio with respect to solar values of Anders \& Grevesse (1989).}
\tablenotetext{b}{The models used in the fits are: Abs($N_{\rm H, Gal}$)*Abs($N_{\rm H}$)\{power-law($\Gamma$) + vmekal($kT_1$, Fe/O)\} and Abs($N_{\rm H, Gal}$)*Abs($N_{\rm H}$)\{vmekal($kT_1$, Fe/O) + mekal($kT_2$)\}, respectively, where Abs is a photo-electric absorption model and (v)mekal is a thermal plasma (with variable Fe abundande).}
\end{deluxetable}

\subsection{Hard X-ray Extranuclear Emission. High-mass X-ray Binaries} \label{HMXBs}

As already mentioned in the previous section, the shape of the hard (2-8 keV) X-ray spectrum in region E is also consistent with a power-law 
energy distribution. This could in principle be explained as due to the contribution of HMXBs. However, arguments based on the absolute
hard X-ray luminosity measured in this region do not favor this hypothesis. An entire galaxy like the Milky Way has about 50 HMXBs (Icken et al.
1995) for an estimated star formation rate of 0.68 to 1.47 M$_{\odot}$ yr$^{-1}$ (Robitaille \& Whitney 2010), with a mean hard X-ray 
luminosity of 5 $\times$ 10$^{37}$ erg s$^{-1}$ per HMXB (Persic et al. 2004). So, in order to explain the hard X-ray luminosity 1.4 $\times$
10$^{40}$ erg s$^{-1}$ emitted by region E, a total of about 280 HMXB, i.e. close to six times higher than the number of HMXBs in the entire Milky Way,
would be required to exist in region E, a small region of less than 400 parsec radius (upper limit given by {\it Chandra} angular resolution). 
Such a high concentration of low luminosity HMXBs is very unlikely.  

NGC 5135 is a luminous infrared galaxy forming stars at a rate much higher than the Milky Way.
It is known that the luminosity function of HMXBs appears to be universal and proportional to the star formation rate of the galaxy 
(Grimm, Gilfanov \& Sunyaev 2003). As a consequence, the luminosity of the brightest HMXB and the number of HMXBs above a certain luminosity increases with the star formation rate in galaxies. This is empirically supported by the non detection in the Milky Way of HMXBs with luminosities above 
2 $\times 10^{38}$ erg s$^{-1}$ (Grimm, Gilfanov \& Sunyaev 2002) and the detection of several more luminous HMXBs in nearby starburst galaxies. In M82, the closest prototype of central starburst galaxy, 22 HMXB with luminosities in the 0.004 to 3 $\times$ 10$^{39}$ erg s$^{-1}$ within the central region of 2.5 kpc $\times$ 2.5 kpc have been detected with {\it Chandra} (Griffiths et al. 2000).   
Moreover, in The Antennae, a well-known nearby pre-coalecense merger with and infrared luminosity a factor of 4 lower than that of NGC 5135, a large number of luminous (49) and highly luminous (11 out of 49) HMXBs have been detected with luminosities above 5 $\times$ 10$^{37}$ erg s$^{-1}$ and 10$^{39}$ erg s$^{-1}$, respectively (Zezas et al. 2002a).  
These luminous HMXBs are predominantly associated with young stellar clusters, i.e. young stellar populations, and distributed over most of
the optical extent of the galaxy (Zezas et al. 2002b), i.e. over a region of about 12 kpc in size.  
So, even if luminous HMXBs as those detected in M82 and The Antennae exist in NGC 5135, a large number (10-100) of these would be required to be concentrated
in a small area in order to explain the hard X-ray luminosity measured in region E. The resulting high density of luminous HMXBs would be orders of magnitude higher than those derived for M82 and The Antennae and unlikely to occur. However, it could still be statistically possible that a single extremely luminous HMXB with a luminosity close to the cut-off 
(2 $\times$ 10$^{40}$ erg s$^{-1}$) of the luminosity function would exist in NGC 5135 and be located in region E. Highly variable ultraluminous X-ray sources (ULXs)  have recently been detected in NGC 1365 (Strateva \& Komossa 2009), and even in M82 (M82 X1;
Voss et al. 2011). If this were the case, a measurable variability of the X-ray emission in this region would be expected (Gilfanov, Grimm \& Sunyaev 2004). A monitoring of NGC 5135 with {\it Chandra} would be needed to confirm or reject this scenario.

\subsection{Hard X-ray Extranuclear Emission. Supernova-driven hot outflowing winds} \label{HXSN}

Detailed X-ray imaging of M82 has shown that the structure of the hard X-ray diffuse emission is similar to that of the nuclear starburst, as traced by the spatial distribution of the compact radio sources, most of which are young supernova remnants (Strickland \& Heckman 2007). Moreover, the extent of the hard X-ray diffuse emission is much smaller than the overall size of the diffuse soft X-ray emission that extends above and below the plane of the galaxy, i.e. minor axis,  up to distances of 5-6 kpc (Strickland \& Heckman 2007, 2009 and references therein).  
This has been interpretated as the unambiguous sign of young massive stars and their associated supernova-driven winds as the direct origin of the diffuse hard X-ray emission.  Similarly in NGC 5135, 
the spatial coincidence of the peak of the [FeII] emission, the
apex of the [FeII] outflow and the secondary 
 X-ray emission peak favors a common physical origin for 
these emissions. The [FeII] emission with its associated outflow has its
natural explanation as the combined effect of supernovae explosions in this
region of the galaxy. Bedregal et al. (2009) derived a supernova rate of 0.05 to 0.1 yr$^{-1}$    
for region E using the radio and [FeII] luminosities and various empirical and model calibrations
(see references in Bedregal et al. 2009) . Since each supernova releases a total of 10$^{51}$ erg 
in kinetic energy (Chevalier 1977), the expected energy release for the estimated supernova 
rate amounts to 1.6 $-$ 3.2 $\times$ 10$^{42}$ erg s$^{-1}$. This value agrees remarkably
well with the predicted mechanical energy in shocks derived from the [FeII] luminosity. 
Assuming an [FeII]1.257$\mu$m/[FeII]1.644$\mu$m line ratio of 1.36 (Nussbaumer \& Storey, 1988), and
typical [FeII]1.257$\mu$m luminosity to shock energy ratios of 3-5 $\times$ 10$^{-3}$ for 
velocity shocks of 75-150 km s$^{-1}$ (Mouri et al. 2000), a total energy in shocks 
equal to 2.7 $\times$ 10$^{42}$ erg s$^{-1}$ is obtained. 

An additional estimate
of the mechanical energy can also be derived directly from the outflowing velocity
and the shocked gas mass associated with the [FeII] emission.
The overall kinetic energy is given as 0.5 $\times$ M$_{shocks}^{gas}$ $\times$ V$^2_{outflow}$,
where the mass of gas is directly proportional to the [FeII] luminosity and inversely proportional to
the particle density of the gas.
Assuming solar abundances, an [FeII] emissivity per unit of volume of
1.3 $-$ 1.6 $\times$ 10$^{-20}$ ergs cm$^3$ s$^{-1}$ (Mouri et al. 2000), particle densities of about 10$^3$ 
cm$^{-3}$ (i.e. well below the critical density of the [FeII] lines), and intrinsic outflowing velocities of about 
100 km s$^{-1}$, the derived kinetic energy is $\sim$ 1 $\times$ 10$^{52}$ ergs. This would be equivalent to the 
combined kinetic energy release of between 10 and 100 supernovae for a high to moderate thermalization efficiency, respectively.

The strong [FeII] emitting region (E in figure 1), although spatially resolved in the VLT/SINFONI observations, is compact with an upper limit to its effective (i.e. half-light) radius (R$_e$) of 80 pc. Therefore, the [FeII] emission detected in this region is consistent with the combined effect of a large
number of supernovae confined in a small volume. A high number of supernovae have also been identified at radio frequencies in the nuclear region (150 pc in size) of  Arp299A (IC 694), a well known LIRG (Perez-Torres et al. 2009; Bondi et al. 2012). Under these conditions, models developed to explain the diffuse hard X-ray emission in
M82 (Strickland \& Heckman 2009) predict that an important fraction of the kinetic energy 
released by the supernovae will actually drive bulk motions of the surrounding interstellar gas, i.e. will have a high 
thermalization efficiency. As a consequence, the surrounding ISM will increase its temperature to tens to hundreds 
of million Kelvin, emitting not only soft but also hard X-rays. 
In these models, for a given thermalization efficiency ($\epsilon$) and mass loading factor ($\beta$), the hard X-ray luminosity is inversely proportional to the size of the region
(L$_{HX} \propto \epsilon^{-1} \times \beta^3 \times R_{e}^{-1}$;  see equations (8) \& (16) in  Strickland \& Heckman 2009).
A 10 Myr old instantaneous burst
releasing a total (supernova and massive stars) mechanical energy of 4.5 $\times$ 10$^{42}$ erg s$^{-1}$ with a high thermalization efficiency
($\epsilon=$ 1.0) and no mass loading ($\beta$= 1.0) in a region of 285 pc radius would 
produce a hard X-ray luminosity of 5 $\times$ 10$^{38}$ erg s$^{-1}$ (model C, Strickland \& Heckman 2009). 
Scaling linearly with the 
effective radius of the [FeII] emitting region (R$_e$ $\leq$ 80 pc), the expected hard X-ray luminosity for region E would be
about 2 $\times$ 10$^{39}$ erg s$^{-1}$, a factor of seven less than measured (see Table 1). This predicted This predicted hard X-ray luminosity is however a lower limit. It could largely increase if the thermalization efficiency were lower than 1 and in particular, if the  
mass loading factor, that represents the contribution to the total mass swept by the wind due to the cold ambient interstellar medium, were larger than 1 
(L$_{HX} \propto \beta^3$). The measured properties of the diffuse hard X-ray emission in M82
are compatible with a range of models characterized by thermalization efficiencies between 0.3 and 1.0, and
mass loading factors between 1.0 and 2.8 (Strickland \& Heckman 2009).  In the highly active star-forming and dense molecular gas environments of the central regions of LIRGs, similar conditions as those in M82 are likely to be present. Therefore, the measured hard X-ray luminosity in NGC 5135 would be consistent with the predictions for a mass loading factor and thermalization efficiency of 2.5 and 0.5, respectively. In addition, the total luminosity predicted by the models as due to the hot thermalized medium is about 2.5 times higher than its hard X-ray luminosity (see Table 7 of Strickland \& Heckman 2009), i.e. close to the 4.5 factor measured for the soft to hard X-ray emission ratio in region E.

\subsection{Hard X-ray Emission and Star Formation Rates} \label{XRSFR}

As in many other wavelengths (UV, H$\alpha$, infrared, radio), the X-ray 
luminosity has also been used as a tracer of the star formation in starburst galaxies. Empirical
linear relations have been established between soft and hard X-ray luminosities and star formation rates (SFR), under the
assumption of constant star formation (e.g. Pereira-Santaella et al. 2011, and references therein). These relations
are obtained using the observed infrared continuum and optical recombination line luminosities 
as direct tracers of the obscured and unobscured star formation rates. It is also generally accepted that in the absence of an AGN, the 
hard X-ray emission in galaxies is due to low- and high-mass X-ray binaries (LMXB and HMXB, respectively), with HMXB dominating 
the hard X-ray emission in extreme 
starbursts such as non-AGN LIRGs and ULIRGs (Persic \& Rephaeli 2002, Grimm et al. 2003, 
Lehmer et al. 2010). 

Although these relations appear to hold for starbursts covering a wide range of SFRs, they show a factor of four peak-to-peak scatter (one sigma) around the
best fit (e.g Pereira-Santaella et al. 2011). This scatter is likely to have a real physical origin and is worth discussing further. On the one hand, the empirical data used to derive the obscured (i.e. infrared bright) and unobscured (i.e. UV or optical
bright) star formation lacks usually the appropriate spatial resolution and assumes implicitly that these regions also emit hard X-rays. This is
clearly not the case for NGC 5135 where, outside the nucleus, the dominant Pa$\alpha$ (Alonso-Herrero et al. 2006) and Br$\gamma$ (Figure 1) 
emission peaks correspond to a circumnuclear region which is spatially independent from the dominant hard X-ray emitting region and located at a distance of about 1.2 kpc from it. Moreover, subarcsec mid-IR imaging (Alonso-Herrero et al. 2006) shows that the mid-IR emission outside the AGN is distributed in four almost equally bright regions, the Br$\gamma$ and hard X-ray peaks and two other regions at distances of about 1 kpc from the hard X-ray emitting peak. 
On the other hand, the assumption that the X-ray emission is dominated by the contribution of compact sources associated with LMXB and HMXB is not 
necessary realistic. Evidence for the presence of diffuse, extended soft and hard X-ray emitting regions has now been firmly established
in the nuclear starburst region of nearby galaxies like NGC 253 
(Weaver et al. 2002, Mitsuishi, Yamasaki \& Takei 2011), and M82 (Strickland \& Heckman 2007), although with different physical origins. The hard X-ray emission in NGC 253 appears to be due not only to photoionization by 
 a low luminosity AGN (Weaver et al. 2002) but also associated with the combined energy output of tens to hundreds of supernovae (Mitsuishi et al. 2011). The extent of the hard X-ray emission in M82 is similar to
the distribution of the radio sources (e.g. young supernova remnants). Therefore it is likely that in NGC 253 and M82 there is a 
relation between 
supernova explosions, and subsequent winds, and the diffuse hard X-ray emitting gas (Strickland \& Heckman 2007). 
As already mentioned ($\S$4.3), detailed modeling of the supernova-related superwinds
predicts the generation of diffuse hard X-ray emitting plasma in compact and highly thermalized starburst regions (Strickland \& Heckman 2009). 

Taking M82 as a prototype of a low luminosity starburst (i.e. L$_{IR} < 10^{11} L_{\odot}$), the diffuse component
is responsible for most (90\%) of its soft X-ray emission in the nuclear region (0.5 kpc radius). In hard X-rays, while the dominant contribution is due 
to point sources likely associated with binaries (Strickland \& Heckman 2007), the diffuse component makes a relevant
contribution ($\sim$20\% to 30\%) to the total hard X-rays emission (see Table 2 of Strickland \& Heckman 2007). Moreover, recent X-rays studies of a
large sample of luminous and ultraluminous infrared galaxies (log L$_{IR}$=  11.73 -12.57 $L_{\odot}$) show in non-AGN sources an X-ray excess compatible with 
strong Fe XXV emission at 6.7 keV (Iwasawa et al. 2009). The presence of this high-ionization Fe line is incompatible with HMXB emission and it
is likely due to a high-temperature ($\it T \sim 10^8$ K) plasma (Iwasawa et al. 2009). These authors also conclude that unlike for local star-forming galaxies, HMXBs are not the primary source of the hard 
X-ray emission seen in these non-AGN U/LIRGs, suggesting that a transition in the nature of the dominant hard X-ray source from massive 
X-ray binaries to hot diffuse plasma occurs somewhere in the range log (L$_{IR}$/L$_{\odot}$)= 11.0-11.73 (Iwasawa et al. 2011). Even more, recent X-ray studies of the nuclear region of NGC 253 reinforce this scenario. These studies (Mitsuishi et al. 2011) demonstrate that the FeK line 
complex emission is extended and distributed in a 60 arcsec$^2$ region around the nucleus, with the highly ionized lines explained by the cumulative emission of hundreds of supernova with a minor contribution from binaries.  

NGC 5135 with an infrared luminosity (L$_{IR}$) of 2 $\times 10^{11} L_{\odot}$ can be considered as a prototype of LIRG that happens to be close enough for a detailed investigation of the spatial distribution and physical origin of the hard X-ray emission. As already shown
(see $\S$ 3), the hard X-ray luminosity is dominated (82\%) by the AGN, while the rest is associated with the [FeII] peak (9\%) and regions
confined within 1 kpc around it (9\%). So, aside from the AGN contribution, the hard X-ray emission in NGC 5135 is due to supernova-driven hot plasma confined in small regions. This physical scenario could also be present in an important fraction of non-AGN LIRGs and ULIRGs. 
The overall size of the starburst region in LIRGs traced by near-IR and optical hydrogen lines (Alonso-Herrero et al. 2006a, Rodriguez-Zaurin et al. 2011), or mid-IR continuum emission (Arribas et al. in preparation) is typically few kpc, distributed in few to several high surface brightness clumps of hundreds of pc, each. Our ongoing SINFONI survey (Piqueras et al. in preparation) shows that many nearby LIRGs have few to several bright compact [FeII] emitting regions in their (circum)nuclear regions.  Moreover, some of these LIRGs (NGC 3256 and IC 694) do show in their integrated X-ray spectrum highly-ionized iron K$\alpha$ emission lines (Pereira-Santaella et al. 2011, Ballo et al. 2004), supporting therefore the hypothesis that an important, if not dominant, fraction of the hard X-rays in these galaxies is associated to hot plasma emission. 

If the hard X-ray emission in non-AGN U/LIRGs is not dominated by the HMXB emission but rather by a supernova-driven hot X-ray emitting plasma, the meaning of the star formation rates (SFR) derived from the hard X-ray luminosity have to be reconsidered. The X-rays emission of HMXB and hot plasma are due to different physical mechanisms, accretion versus thermalized hot plasma, and subject to different physical parameters. In fact, while the number of luminous HMXBs appears to scale linearly with the SFR (Persic et la. 2004), the hot X-ray emitting gas shows a non-linear relation with the SFR according to models 
(L$_{HX}$$\propto SFR^2$; Strickland \& Heckman 2009). Moreover, while both supernova and HMXB 
are associated with massive stars, their birthrate and lifespan differ significantly
(Persic \& Rephaeli 2002; Iben et al. 1995), and therefore the associated star formation rate would also be different. An indication that this could be the case is given by the discrepancy in the derived SFR values for NGC 5135 when calculated using the different empirical
relations based on the 24 $\micron$, soft and hard X-ray luminosities (see Pereira-Santaella et al. 2011 for specific relations).
NGC 5135 has an AGN and therefore corrections to its contribution to the luminosity of the galaxy have to be applied before
calculating SFRs. The AGN contribution to the 24 $\mu$m luminosity was derived from the decomposition of the Spitzer Infrared Spectrograph low-resolution 5-38 $\mu$m spectrum into AGN and starburst components using clumpy torus models and star-forming galaxy templates (Alonso-Herrero et al. 2012). The AGN soft and hard X-ray luminosity were derived directly from the {\it Chandra} images (see Table 1 for specific values), and subtracted from the integrated X-ray emission.  
Correcting for the AGN as above, the total SFR of the central 2 kpc starburst derived from the hard X-ray luminosity, gives a lower limit with  a factor of
2 and 4 lower than that obtained from 
the infrared and soft X-rays luminosities, respectively. A plausible explanation for this discrepancy is that the regions dominating the integrated emission at different wavelengths appear at different locations within the galaxy and separated by hundreds of parsecs, i.e. different physical origin, 
different evolutionary phases and no causal connection between the regions. This is the case for NGC 5135 where,
aside from the AGN, the hard X-rays emission appears to be dominated ($> 50\%$) by region E, the soft X-rays shows a more extended emission with contributions from region E, as well as other compact regions and diffuse emission (Figure 1 and Levenson 
et al. 2004), and the mid-IR emission is distributed in several circumnuclear star-forming regions located at distances of about 1-1.2 kpc from
region E (Alonso-Herrero et al. 2006b). This is in agreement with a scenario where for a given luminous infrared galaxy, the hard X-rays would be produced mainly in a subset of star-forming regions characterized by its compactness (i.e. less than hundred parsecs), very high SFR densities (L$_{HX} \propto SFR^2/R$ according to Strickland \& Heckman 2009 models) and a large supernova rate, while the soft X-rays and IR emission would trace the young massive stars associated with the overall star formation in larger (kpc size) regions. Under this scenario, the hard X-rays emission would give a lower limit to the
SFR, as is the case in NGC 5135. This scenario would also give a natural explanation other than the absorption effects, to the deficit of hard X-ray emission measured in LIRGs and ULIRGs with respect to the values expected from their IR-derived star formation rates (Lehmer et al. 2010).

High-angular X-ray and [FeII] imaging of a large sample of nearby LIRGs are needed in order to further
investigate the origin, and quantify the contribution, of starburst-driven hot ISM to the hard X-ray emission in luminous infrared galaxies, and therefore
validate the use of hard X-ray luminosity to derive accurate estimates of star forming rates in these galaxies.

\section{Conclusions}
Near-infrared integral field spectroscopy of the luminous infrared galaxy NGC 5135 combined with archival
high resolution {\it Chandra} X-ray imaging has been used to investigate in detail the physical origin of its hard X-ray emission
and to establish the validity of the hypothesis behind the star formation rates derived from the hard X-ray luminosities in
luminous and ultraluminous infrared galaxies. The main conclusions are:

\begin{itemize}

\item The AGN is identified in the near-IR by a strong [SiVI] 1.96$\mu$m emission showing a nuclear component and
  an extended diffuse elongation of about 0.6 kpc towards the north of the nucleus. The AGN generates 90\% of the hard X-ray luminosity emitted within 
the central 2 kpc diameter. There are hints of a connection between the diffuse [SiVI] emitting gas and the hard X-ray emission north of the AGN. 

\item The strongest [FeII] emission peak (region E) is located at a distance of about 0.9 kpc southwest of the AGN  showing a local velocity field
 with the kinematical characteristics of outflowing material. The [FeII] luminosity and kinetic energy in this region is consistent with the
 scenario of supernova-induced outflows and with previous estimates of the supernovae rate (0.05 to 0.1 SN yr$^{-1}$) in this region.

\item The luminous hard X-ray emitting region outside the AGN coincides with the strongest [FeII] peak and associated outflowing material. 
A direct causal connection appears to exist between supernovae-induced winds and shocks, and the hard X-ray emitting ISM in this region. This result is
in agreement with recent scenarios where starbursts confined in small volumes and with high thermalization efficiencies are able to
heat the ISM to temperatures of 10 to 100 million degrees, emitting therefore hard X-rays. Starbursts in U/LIRGs are compact and confined to small 
regios and therefore the hot diffuse X-ray emitting gas could represent an important (even dominant) contribution
to the overall hard X-ray emission.

\item The alternative explanation of the hard X-ray emission  outside the AGN as due to HMXBs is not favoured although cannot be ruled out. If luminous
HMXBs similar to those detected in other nearby starburst galaxies like M82 and The Antennae exist in NGC 5135, a much higher density of HMXBs ($>$ 10-100
pc$^{-2}$)  would be required to explain the hard X-ray luminosity in region E. The possibility of having a single, extremely luminous HMXB with a luminosity close to the cut-off limit (L$_{HX}$= 2 $\times$ 10$^{40}$ erg s$^{-1}$) of the HMXB luminosity function is still compatible with the measured luminosity. 

\item The SFR in NGC 5135 derived from the hard X-ray luminosity gives a lower limit corresponding to factors 2 and 4 lower than the 24 $\mu$m and soft X-ray derived values, respectively. This is understood as a consequence of the contribution of the different star-forming regions. The dominant mid-IR (and Br$\gamma$) emitting regions are spatially separated by distances of about 1 kpc from the extended soft and compact hard X-ray emitting regions. Thus, the individual star-forming regions are physically disconnected and appear to be in different evolutionary states (i.e. age, ISM), contributing differently to the emission at various wavelengths.  

\item The origin and nature of the hard X-ray emission in luminous infrared galaxies like NGC 5135 has to be investigated in more detail through high angular resolution X-ray, complemented with near- and mid-infrared imaging, in a larger sample of galaxies. If, according to the models and the evidence shown so far in 
NGC 5135, the hard X-ray emission trace the more compact star-forming regions with the highest star-forming densities 
(L$_{\rm HX} \propto SFR^2/R$), the hard X-rays emission in these galaxies would appear associated with compact, luminous [FeII] emitting star-forming regions, 
and would represent a lower limit to the overall SFR, traced by the more extended IR and soft X-ray emission.

\end{itemize}

\acknowledgments

MP-S acknowledges support from CSIC under grant JAE-Predoc-2007. This work
has been supported by the Spanish Plan Nacional del Espacio under grants
ESP2007-65475-C02-01 and AYA2010-21161-C02-01.


\begin{thebibliography}{}
\bibitem[Alonso-Herrero et al.(2003)]{AAH03} Alonso-Herrero, A. et al., 2003, \aj, 125, 1210
\bibitem[Alonso-Herrero et al.(2006a)]{AAH06a} Alonso-Herrero, A. et al., 2006a, \apj, 650, 835
\bibitem[Alonso-Herrero et al.(2006b)]{AAH06b} Alonso-Herrero, A. et al., 2006b, \apj, 652, L83
\bibitem[Alonso-Herrero et al. (2012)]{AAH12} Alonso-Herrero, A., Pereira-Santaella, M., Rieke, G.H., \& Rigopoulou, D.
2012, \aj, 744, 2
\bibitem[Arnaud (1996)]{Arnaud96} Arnaud, K. A. 1996, Astronomical Data Analysis Software and Systems V, 101, 17
\bibitem[Bedregal et al. (2009)]{Bed09} Bedregal, A. et al. 2009, \apj, 698, 1852
\bibitem[Bedregal et al. (2012)]{Bed11} Bedregal, A. et al. 2012, in preparation
\bibitem[Bondi et al. (2012)]{Bon12} Bondi, M., Perez-Torres, M.A., Herrero-Illana, R., Alberdi, A. \aap, (in press) (arXiv 1201.3220)
\bibitem[Chevalier (1977)]{Che77} Chevalier, R.A. 1997, \araa, 15, 175
\bibitem[Clements et al. (2002)]{Cle02} Clements, D. L. et al. 2002, \apj, 581, 974
\bibitem[Colina (1993)]{Col93} Colina, L., 1993, \apj, 411, 565
\bibitem[Genzel et al. (1998)]{Genzel98} Genzel, R., et al. 1998, \apj, 498, 579
\bibitem[Gilfanov et al. (2004)]{Gilfanov04} Gilfanov, M., Grimm, H.J., \& Sunyaev, R. 2004, \mnras, 351, 1365 
\bibitem[Greenhouse et al. (1991)]{Greenh91} Greenhouse, M. 1991, \apj, 383, 164
\bibitem[Griffiths et al. (2000)]{Grif00} Griffths, R.E., 2000, Science, 290, 1325
\bibitem[Grimes et al. (2005)]{GHSP05} Grimes, J. P., Heckman, T., Strickland, D., \& Ptak, A. 2005, \apj, 628, 187
\bibitem[Grimm et al. (2002)]{Grimm02} Grimm, H.J., Gilfanov, M., \& Sunyaef, R. 2002, \aap, 391, 923
\bibitem[Grimm et al. (2003)]{Grimm03} Grimm, H.J., Gilfanov, M., \& Sunyaef, R. 2003, \mnras, 339, 793 
\bibitem[Gonzalez-Delgado et al. (1998)]{GonD98} Gonzalez-Delgado, R. et al. 1998, \apj, 505, 174
\bibitem[Iben et al. (1995)]{Iben95} Iben, I., Tutukov, A.V., \& Yungelson, L.R. \apss, 100, 217
\bibitem[Iwasawa et al. (2009)]{Iwa09} Iwasawa, K. et al. 2009, \apj, 695, L103
\bibitem[Iwasawa et al. (2011)]{Iwa11} Iwasawa, K. et al. 2011, \aap, 529, 106
\bibitem[Lehmer et al. (2010)]{Leh10} Lehmer, B.D. et al. 2010, \apj, 724, 559
\bibitem[Levenson et al. (2004)]{Lev04} Levenson, N.A. et al. 2004, \apj, 602, 135
\bibitem[Mcdowell et al. (2003)]{Mcd03} McDowell, J.C. et al. 2003, \apj, 591, 154
\bibitem[Mitsuishi et al. (2011)]{MYT11} Mitsuishi, I., Yanasaki, N.Y., \& Takei, Y. 2011, \apjl, 742, L31
\bibitem[Mouri et al. (2000)]{Mou00} Mouri, H., Kawara, K., \& Taniguchi, Y. 2000, \apj, 528, 186
\bibitem[Nandra et al. (2007)]{Nan07} Nandra, K., et al. 2007, \mnras, 382, 194
\bibitem[Nardini et al. (2008)]{Nard08} Nardini, E., et al. 2008, \mnras, 385, L130
\bibitem[Nussbaumer \& Storey (1988)]{Nuss88} Nussbaumer, H., \& Storey, P.J. 1988, \aap, 193, 327
\bibitem[Pereira-Santaella et al. (2011)]{PS11} Pereira-Santaella, M. et al. 2011, \aap, 535, A93 \\
\bibitem[Perez-Torres te al. (2009)]{PT09} Perez-Torres, M., Romero-Ca\~nizales, C., Alberdi, A., \& Polatidis, A. 2009, \aap, 507, L17 
\bibitem[Persic \& Rephaeli (2002)]{Per02} Persic, M., \& Rephaeli, Y. 2002, \aap, 382, 843
\bibitem[Persic et al. (2004)]{Per04} Persic, M. 2004, \aap, 419, 849
\bibitem[Risaliti et al. (2000)]{Ris00} Risaliti, G. et al., \aap, 357, 13
\bibitem[Rodriguez-Zaurin et al. (2011)]{RZ11} Rodriguez-Zaurin, J. 2011, \aap, 527, 60 
\bibitem[Singh et al. (2011)]{SRBSh11} Singh, V et al. 2011, \mnras, in press (arXiv1109.4342)
\bibitem[Strateva \& Komossa (2011)]{StK11} Strateva, I.K., \& Komossa, S. 2009, \apj, 692, 443
\bibitem[Strickland et al. (2002)]{Strickland02} Strickland, D. et al. 2002, \apj, 568, 689
\bibitem[Strickland et al. (2004)]{Strickland04} Strickland, D. et al. 2004, \apjs, 151, 193
\bibitem[Strickland \& Heckman (2007)]{Strickland07} Strickland, D. \& Heckman, T.M. 2007, \apj, 658, 258
\bibitem[Strickland \& Heckman (2009)]{Strickland09} Strickland, D. \& Heckman, T.M. 2009, \apj, 697, 2030
\bibitem[Vanzi \& Rieke (1997)]{Vanzi97} Vanzi, L. \& Rieke, G.H. 1997, \apj, 479, 694
\bibitem[Voss et al. (2011)]{Voss11} Voss, R. et al. 2011, \mnras, 418, L124
\bibitem[Weaver et al. (2002)]{Weaver02} Weaver, K.A. 2002, \apj, 576, L19
\bibitem[Wilms et al. (2000)]{WAMc00} Wilms, J., Allen, A., \& McCray, R. 2000, \apj, 542, 914
\bibitem[Zezas et al. (2002a)]{Zezas02a} Zezas, A., Fabbiano, G., Rots, A.H., \& Murray, S.S. 2002a, \apss, 142, 239
\bibitem[Zezas et al. (2002b)]{Zezas02b} Zezas, A., Fabbiano, G., Rots, A.H., \& Murray, S.S. 2002b, \apj, 577, 710
\end{thebibliography}
\end{document}